# Letter

# A dust-parallax distance of 19 megaparsecs to the supermassive black hole in NGC 4151

Sebastian F. Hönig[1,2], Darach Watson[1], Makoto Kishimoto[3], Jens Hjorth[1]

**The active galaxy NGC 4151 has a crucial role as one of only two active galactic nuclei for which black hole mass measurements based on emission line reverberation mapping can be calibrated against other dynamical methods[1-3]. Unfortunately, effective calibration requires an accurate distance to NGC 4151, which is currently not available[4]. Recently reported distances range from 4 to 29 megaparsecs (Mpc)[5-7]. Strong peculiar motions make a redshift-based distance very uncertain, and the geometry of the galaxy and its nucleus prohibit accurate measurements using other techniques. Here we report a dust-parallax distance to NGC 4151 of $D_A = 19.0^{+2.4}_{-2.6}$ Mpc. The measurement is based on an adaptation of a geometric method proposed previously using the emission line regions of active galaxies[8]. Since this region is too small for current imaging capabilities, we use instead the ratio of the physical-to-angular sizes of the more extended hot dust emission[9] as determined from time-delays[10] and infrared interferometry[11-14]. This new distance leads to an approximately 1.4-fold increase in the dynamical black hole mass, implying a corresponding correction to emission line reverberation masses of black holes if they are calibrated against the two objects with additional dynamical masses.**

The central black hole in AGN is surrounded by a putative accretion disk that emits pre-dominantly at ultraviolet and optical wavelengths. At large distances from this central emission source, the gas is cool enough for dust to survive (temperature ≤1500 K). This "dusty torus" absorbs the UV/optical radiation and thermally reemits the energy in the infrared (IR). Thus, any variability in the UV/optical emission will be detected in the dust emission with some time delay. Near-IR reverberation mapping measures the time lag, $\tau$, between the UV/optical variability and the corresponding changes in emission of the hot dust, which is located about the sublimation radius $R_{sub} = R(T \sim 1,500$ K). The time lag can be converted into a physical size via $R_\tau = \tau \cdot c$, where $c$ is the speed of light. Typically, time lags are found in the range of several 10s to 100s of days, which corresponds to physical sizes of the order of 0.1 pc, with a square-root dependence on luminosity[12, 15].

In parallel, IR interferometry taken at the same wavelength measures a corresponding angular size, $\rho$, of the same emission region. The angular and physical sizes are trigonometrically related by $\sin \rho = R_\tau/D_A$, where $D_A$ is the angular-diameter distance to the object. For small angles $\sin \rho \approx \rho$ and accounting for cosmological time dilation, we obtain $D_A(\mathrm{Mpc}) = 0.173 \times \tau(\mathrm{days}) / (\rho(\mathrm{mas}) \cdot (1+z))$, which forms the basis of the distance measurement presented here (see Fig. 1). A geometric technique was first proposed for broad emission lines[8]. Unfortunately, the typical angular size of the broad-emission line region for bright AGN is of the order of 0.001 – 0.01 milliarcseconds, which is too small to be spatially resolved with current optical long-baseline interferometers. The dust continuum emission, on the other hand, is larger by a factor of ~4 and IR interferometers have now managed to resolve about a dozen AGN[11,12,14,16]. Moreover, using dust emission requires only photometric reverberation mapping instead of spectrally resolving emission lines. Finally, dust physics is arguably easier to model than gas line emission.

To determine the distance to the supermassive black hole in NGC 4151, we make use of interferometry obtained with the two Keck telescopes and monitoring data from literature. $V$- (wavelength 0.55 μm) and $K$-band (2.2 μm) photometric monitoring from 2001 to 2006[10] trace the UV/optical and hot dust emission, respectively. Since long-term brightness changes can cause $\tau$ and $\rho$ to increase or decrease, monitoring and interferometry data should be taken more or less contemporaneously. We use six Keck interferometry measurements taken between 2003 and 2010[11-14]. These overlap with the monitoring data, and inspection of long-term brightness trends showed that fluctuations were moderate between 2000 and 2010[16]. Indeed, no significant change in size has been detected for this set of interferometry and variability data[16,18].

When comparing angular and physical sizes, it is important to make sure they refer to the same physical region. First, observations of the dust emission are preferably made at the same waveband (here: $K$-band).

[1]Dark Cosmology Centre, Niels Bohr Institute, University of Copenhagen, Juliane Maries Vej 30, 2100 Copenhagen Ø, Denmark. [2]School of Physics & Astronomy, University of Southampton, Southampton SO17 1BJ, United Kingdom. [3]Department of Physics, Faculty of Science, Kyoto Sangyo University, Kamigamo-motoyama, Kita-ku, Kyoto 603-8555, Japan

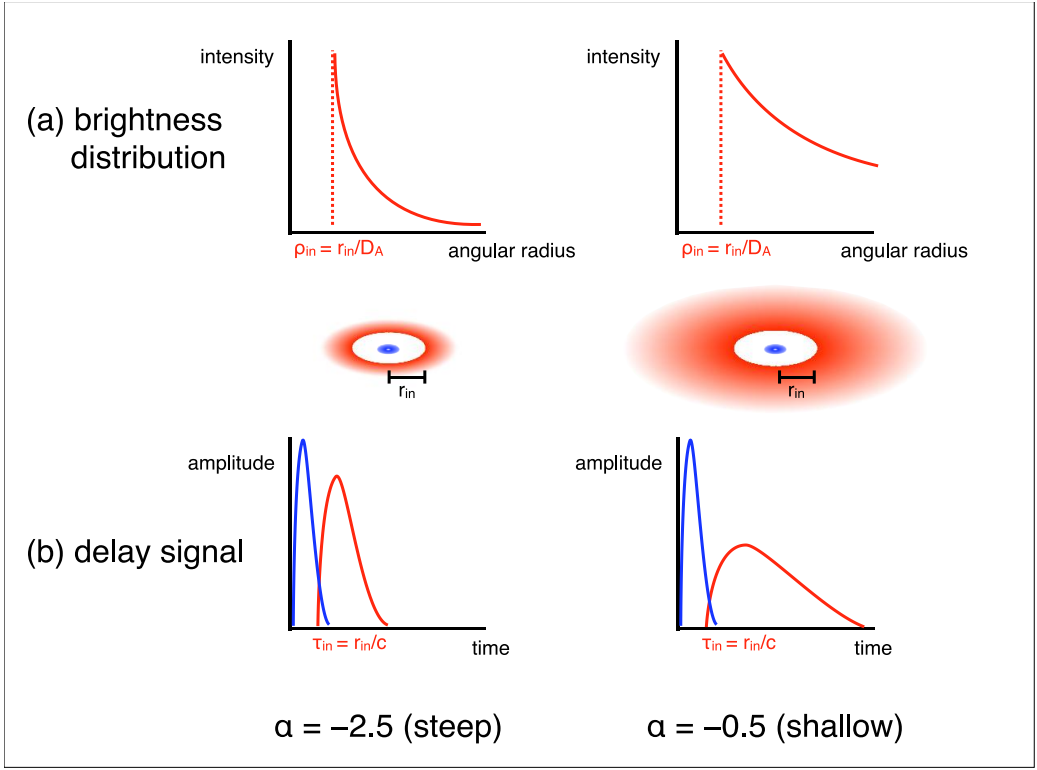

**Figure 1 | Impact of the brightness distribution on the observed sizes and time lags.** The bottom row (b) illustrates effects of varying the brightness distribution on the time lag signal (blue: optical; red: near-IR), while the upper row (a) outlines the corresponding (interferometric) brightness distribution in the near-IR. Compact distributions lead to shorter time lags and smaller interferometric radii than shallow profiles. This information is encoded in the shape (width, amplitude) of the light curve. With a simple power law parametrization, this smearing effect can be accounted for to determine the time lag and angular size of the innermost radius of the brightness distribution. The simultaneous modeling of both light curves and interferometry results in a very precise angular distance measurement.

Second, the spatial distribution of the dust around the AGN impacts the observed sizes (see Fig. 1): Dust that is homogeneously distributed will result in a larger apparent size than a compact dusty region, since the former involves a larger region that contributes to the emission at a given wavelength. As a consequence, the variability signal of the $K$-band will show some degree of smoothening with respect to the $V$-band signal, depending on the distribution. This also involves a shift of the peak time lag. At the same time the size measured by interferometry will appear correspondingly smaller or larger. As shown in literature, this distribution effect in both types of data can be effectively modeled by means of a disk model[17,18], assuming that the dust is heated by AGN radiation and the projected brightness distribution is represented by a power law $S(r) \propto r^{\alpha}$ (see Methods for details). This geometry is in line with theoretical expectations and observational evidence of the hot-dust region[19,20]. Indeed, the model has been successfully applied to reproduce multi-wavelength, multi-baseline interferometry of several AGN (including NGC 4151)[14,21] as well as the light curve of NGC 4151[18,22].

The common reference size in such a model is the inner radius $r_{in}$ of the brightness distribution. For reverberation mapping and interferometry, this corresponds to a reference time lag $\tau_{in} = r_{in}/c$ and angular size $\rho_{in} \approx r_{in}/D_A$ of the inner boundary of the brightness distribution, respectively. Further parameters that may influence the observationally inferred physical and angular size of $r_{in}$ are the disk geometry of the emission region and the dust properties. For the inclination and disk orientation, we use observational constraints based on the dynamics of the emission line region in NGC 4151 and polarimetry[23–25]. We do not consider the radio jet in this study[26], since the available data does not allow for reliably establishing both position angles and inclinations simultaneously. The absorption efficiency of the dust is implicitly included in our parametrization of the brightness distribution. Moreover, the sublimation temperature does not affect the distance determination since it scales in the same way for the reference angular sizes and the reference time lags.

Since the light curves are sampled with finite and varying gaps between each observation, we simulated 1,250 random, continuous representations of data using the AGN variability pattern derived from the structure function. $\tau_{in}$, $\alpha$, and $\rho_{in}$ were calculated simultaneously given the observationally constrained inclination and disk orientation as priors. This resulted in 1,250 estimates of $D_A$, which are shown in Fig. 2. An important feature of this process is that although determining the reference time lag or the reference angular size individually is quite uncertain, both parameters are strongly correlated with the dust brightness distribution. Thus, the ratio, i.e. $D_A$, can be constrained with much higher precision than the reference time lag or reference angular size individually, if $\tau_{in}$ and $\rho_{in}$ are calculated simultaneously given the inferred α.

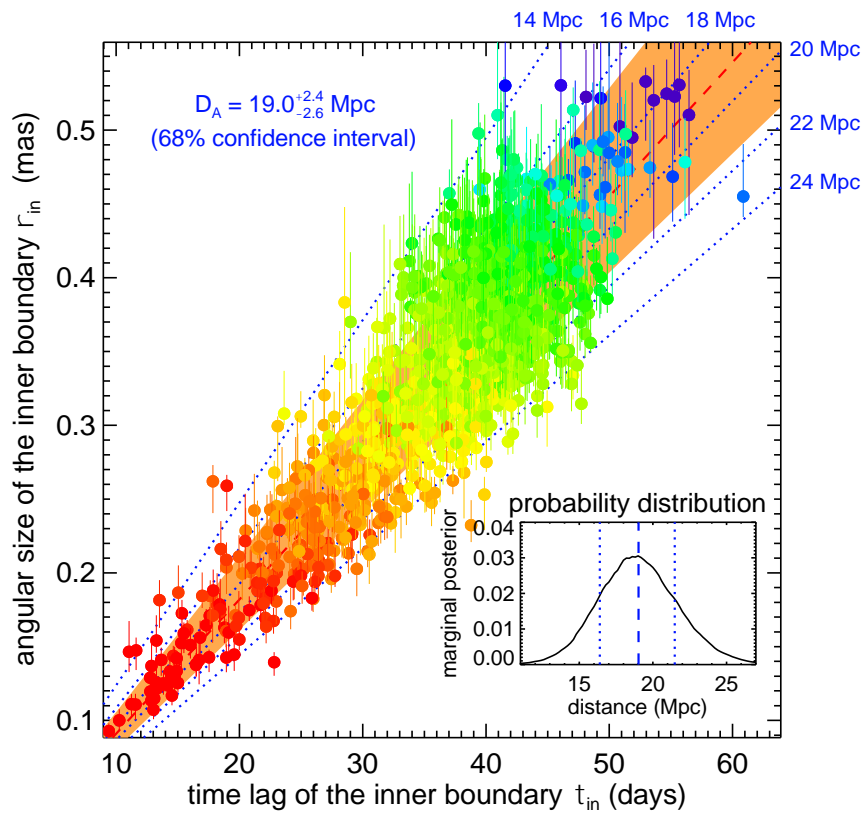

**Figure 2 | Relating time lags and angular sizes to measure the absolute distance to NGC 4151.** The colored circles show the modeled reference time lags $\tau_{in}$ and associated angular sizes $\rho_{in}$ (1,250 random realizations of the V-band light curve; 68% confidence interval shown as error bars). The blue-dashed lines mark $\tau_{in}/\rho_{in}$ ratios corresponding to distance $D_A = 14 - 24$ Mpc. The distribution median and 68% confidence interval are marked by the red-dashed line and orange-shaded area. Brightness distribution power law indices α are color coded from α = 2 (red; shallow) to α = –15 (blue; steep). The small inset shows the probability distribution function of the inferred distance, with mean and 68% confidence interval indicated by blue-dashed and -dotted lines, respectively.

We obtain an angular diameter distance to NGC 4151 of $D_A = 19.0^{+2.4}_{-2.6}$ Mpc (see probability distribution in Fig. 2). The error bars include statistical uncertainties from the reverberation and interferometric observations, as well as the systematic uncertainties introduced by the geometry, brightness distribution, and the uncertainty in host and putative accretion disk contribution to the K-band interferometry. These uncertainties have been accounted for in the Monte Carlo simulations when sampling the data (see Methods). The new distance clarifies the uncertain situation for NGC 4151. The galaxy is in the vicinity of the Virgo cluster (angular separation to the Virgo cluster center <30°), resulting in strong peculiar motion with respect to the Hubble flow[6,27]. Therefore, any recession velocity-dependent distance has to be considered uncertain. Geometric megamaser distances require that the nuclear region is seen very close to edge-on, which is generally not the case for unobscured AGN. Moreover, a direct distance estimate based on the Tully-Fisher relation is difficult because of the face-on view of the galactic disk, which makes it difficult to determine the required rotational velocities. Attempts to estimate the distance this way resulted in a wide range of values between ~4 and 20 Mpc[5,6]. More recently, a luminosity distance of 29.2±0.5 Mpc was suggested based on near-IR reverberation mapping only and a model for absorption and re-emission of dust, which cannot be reconciled with our new result or the other estimates[7].

The new precise distance to NGC 4151 is important since this galaxy is a cornerstone in calibrating black hole masses inferred from different methods[2]; aside from NGC 3227, it is the only galaxy with a suitable mass estimated from reverberation mapping, stellar and gas dynamics. Of both galaxies, NGC 4151 has much better mass constraints[2], so that any systematic offset in distance will almost equally affect the calibration of black hole masses. Dynamical mass estimates relate the rotational velocity field of stars or gas surrounding the black hole to their distances from the AGN. In the process, observed angular distances have to be converted into physical distances for which an absolute distance to the galaxy is required. The most recent mass estimates assume $D_A = 13.2$ Mpc[4,28]. A stellar velocity-based mass was reported as $M^{SD}_{BH} = (3.76 \pm 1.15) \times 10^7$ $M_\odot$[4]. Our new measurement implies that this mass is underestimated by a factor of ~1.4, leading to a revised mass of $M^{SD}_{BH} = (5.4 \pm 1.8) \times 10^7$ $M_\odot$. Similarly, the correction in distance increases the gas-dynamical mass[28] from $M^{GD}_{BH} = 3.0^{+0.75}_{-2.2} \times 10^7$ M$_\odot$ to $M^{GD}_{BH} = 4.3^{+1.2}_{-3.2} \times 10^7$ M$_\odot$.

The new distance and corrected $M_{BH}$ values also affect the correction factor $f$ that has to be invoked when converting reverberation time lags and velocities into black hole masses. The most recent canonical value has been found as $f = 4.31 \pm 1.05$ inferred from comparing reverberation mapping masses to black hole masses based on the $M_{BH}$-$\sigma_*$ relation[29]. Using our corrected values for the dynamical black hole masses on the reverberation data[1], we find a range of $f = 5.2 – 6.5$ (reflecting the difference between gas and stellar dynamical masses), implying a systematic shift to larger masses. Such larger $f$-values may be generally applicable, as also suggested by complex modeling of velocity-resolved reverberation mapping data[30].

**Acknowledgements** We thank Radek Wojtak for helpful discussions on peculiar velocities near the Virgo cluster. S.F.H. acknowledges support from a Marie Curie International Incoming Fellowship within the 7th European Community Framework Programme (PIIF-GA-2013-623804). The Dark Cosmology Centre is funded by the Danish National Research Foundation. This research has made use of the NASA/IPAC Extragalactic Database (NED), which is operated by JPL, Caltech, under contract with the National Aeronautics and Space Administration.

**Author Contributions** S. F. H. and D. W. conceived the project. S. F. H. collected the data, developed the model, and wrote up the paper. D. W. assisted in interpreting the results and helped writing the manuscript. M. K. contributed the interferometry data and help with the data analysis. J. H. contributed to the modeling and interpretation. All authors engaged in discussion and provided comments on the manuscript.

**Author Information** Reprints and permissions information is available at www.nature.com/reprints. The authors declare no competing financial interests. Correspondence and requests for materials should be addressed to S.F.H. (S.Hoenig@soton.ac.uk).

# Methods

**Principle of the measurement**

The present study makes use of the fact that the distance to an object $D_A$ is the ratio between its physical size and the observed angular size. The dust emission in AGN, however, is not a fixed-size object but has emission from a range of radii contributing to any given observed band. Here we focus on the K-band emission for the hottest dust around an AGN. As we previously showed using IR interferometric data[17,21,31], the radius-dependent brightness distribution $F(r)$ in any infrared band can be approximated by a power law

$$F(r) \propto \int Q_{\text{abs};\nu} B_\nu(T(r)) \cdot \left(\frac{r}{r_{\text{in}}}\right)^{\alpha'} \cdot d\nu$$

where $B_\nu(T(r))$ is the frequency-dependent Planck function for temperature $T(r)$ at radius $r$ and $Q_{\text{abs};\nu}$ is the absorption efficiency of the dust. The frequency-integration can be approximated by a power law as $\int Q_{\text{abs};\nu} B_\nu(T) d\nu \propto T^{4+\gamma}$ [32]. The power law index $\gamma$ describes the change of absorption efficiency in the IR with respect to the optical. For black-body radiation, $Q_{\text{abs};\nu} = 1$ and $\gamma = 0$, while typical astronomical dust has $\gamma \approx 1.6 - 1.8$ [33]. Therefore, using the equation of local thermal equilibrium, the emission from radius $r$ can be expressed by a power law[18]

$$F(r) = F(r_{\text{in}}) \cdot \left(\frac{r}{r_{\text{in}}}\right)^{\alpha' - \frac{2}{4+\gamma}} = F(r_{\text{in}}) \cdot \left(\frac{r}{r_{\text{in}}}\right)^{\alpha} \quad (1)$$

where $F(r_{\text{in}})$ is the emission from the innermost radius $r_{\text{in}} = r(T_{\text{sub}})$, where the dust temperature reaches the sublimation temperature $T_{\text{sub}}$. Equation (1) is used to model the observed light curves and interferometry (see below). As such, the dust opacity of the emitting medium is implicitly accounted for since $\gamma$ is absorbed in $\alpha = \alpha' - 2/(4 + \gamma)$, which is directly constrained by observations. Therefore, we do not need to assume any particular dust composition but solve self-consistently for the effect of dust absorption efficiencies.

The rest-frame light-travel time $\tau$ from the AGN to radius r can be expressed as $\tau = r/c$, so that we can rewrite equation (1) as

$$F(\tau) = F(\tau_{\text{in}}) \cdot \left(\frac{\tau}{\tau_{\text{in}}}\right)^{\alpha}. \quad (2)$$

Equivalently, the observed angular size $\rho$ is related to $r$ via $\rho = r/D_A$ (for small $\rho$), so that equation (1) can be written as

$$F(\rho) = F(\rho_{\text{in}}) \cdot \left(\frac{\rho}{\rho_{\text{in}}}\right)^{\alpha}. \quad (3)$$

To measure the angular-diameter distance $D_A$ to the AGN, we have to constrain a time lag and an angular size that are representative of the brightness distribution with power-law index $\alpha$. Conveniently, we choose the scaling time lag $\tau_{\text{in}}$ and $\rho_{\text{in}}$ that correspond to the inner size of the emission region. The power-law index $\alpha$ may be constrained from both interferometry and reverberation measurements. In practice, we do not cover enough baseline lengths with the available interferometry, so that we use the $\alpha$-constraint from fitting the light curves to determine $\rho_{\text{in}}$ (see below for more details).

The absolute values of both scaling sizes $\tau_{\text{in}}$ and $\rho_{\text{in}}$ depend on the choice of $T_{\text{sub}}$. However, according to equations (2) and (3), both $T$-dependencies are exactly the same. Therefore, the ratio of $\tau_{\text{in}}$ and $\rho_{\text{in}}$ is independent of $T$. Nevertheless, we sample $T_{\text{sub}}$ from a normal distribution with mean of 1,400 K and a standard deviation of 100 K. This choice is based on our previous work, where we were simultaneously fitting the near-IR SED and interferometry of NGC 4151 by means of a power-law brightness model[17]. We also tested and confirmed that the distance measurement is, indeed, not depending on temperature. Extended Data Figure 1 shows the $D_A$-$T_{\text{sub}}$ plane of our 1,250 samples. The Spearman correlation coefficient of 0.01±0.03 is very close to zero, implying that the distribution is consistent with being drawn from an uncorrelated parent distribution.

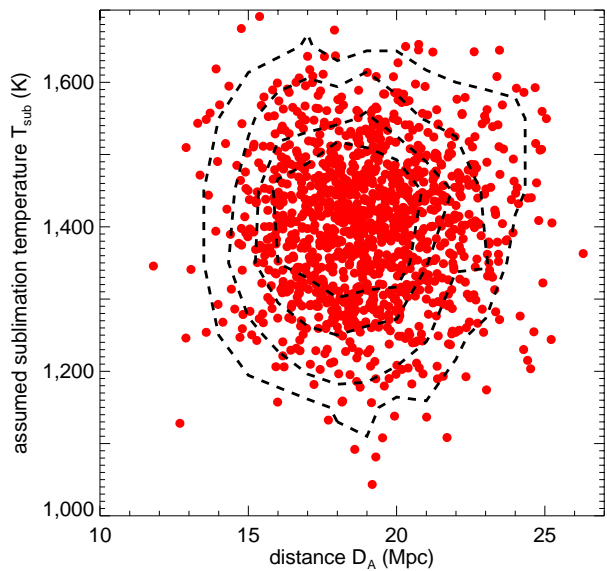

**Extended Data Figure 1 | Dependence of the measured angular diameter distance $D_A$ on the dust sublimation temperature $T_{\text{sub}}$.** The red circles represent the distribution of sublimation temperatures in the 1,250 Monte Carlo runs to model the light curve. Overplotted are dashed contours at 5%, 15%, 33%, and 50% peak density. The distribution of $D_A$ is consistent with being independent of $T_{\text{sub}}$, illustrating that the distance determination is insensitive of the detailed dust properties.

**Data selection and AGN variability**

NGC 4151 has been extensively monitored in the *V*- and *K*-band using the MAGNUM telescope in the period between 2001 and 2006[10]. In 2003, NGC 4151 became the first AGN to be spatially resolved with IR interferometry using the Keck interferometer[11]. Further Keck interferometry were taken in 2009, 2010, and 2011[12,14].

Changes in the luminosity of the putative accretion disk might influence the size of the hot dust region via increased or suppressed sublimation. It has been demonstrated, however, that the *V*-to-*K*-band time lag in NGC 4151 between 2001 and 2006 did not vary significantly despite a peak-to-valley change in luminosity of a factor of ∼20 in the *V*-band, which could have potentially changed the radius by a factor of 4 − 5[18]. Evidence has been found[16] that any change in the size of the hot dust-emitting region in NGC 4151 requires much stronger variability in terms of luminosity and has to be sustained over some time. Furthermore, structural changes appear to be delayed on time scales ∼40 times longer than the time lag (∼5 years)[16]. Based on separate modeling of the light curve and interferometry, respectively, we previously concluded that the inner radius remained unchanged within measurement uncertainties between 2003 and 2010[16,18]. Therefore, the six independent *K*-band interferometry data sets from 2003, 2009, and 2010 can be considered contemporaneous with the photometric monitoring and we include them for our analysis.

**Inclination and orientation of the dusty disk**

The model that is used to recover the time lag $\tau_{in}$ and angular size $\rho_{in}$ of the inner radius of the *K*-band brightness distribution accounts for the geometry of the emitting region in terms of a (projected) disk configuration. We showed previously that it is a simplified, yet accurate, representation of the direct (unobscured) hot dust emission of more complex clumpy or smooth torus models[31].

The position angle of the AGN system/polar axis (=direction of the minor axis of the projected ellipse) has been inferred from kinematic modeling of emission lines and optical polarimetry (see Extended Data Table 1 for a compilation of data). The various constraints originate from different spatial scales: Optical polarization in this type 1 AGN is a result of scattering of the putative accretion disk emission by gas clouds on the scale of the broad-line region, i.e. about a factor of 4 smaller than the hot dust emission. These data indicate an orientation approximately toward PA 90°[25,34]. The outflow kinematics of the narrow emission line region on arcsecond scales point ap- proximately toward PA 35°[23] or 60°[24], with the two different values presumably originating from a different modeling approach. While the smallest spatial resolutions obtained by polarimetry may arguably best trace the orientation of the AGN, we chose a prior on the position angle of 75° and sample from a normal distribution with a standard deviation of 15° to reflect the range of inferred directions. It is important to note that with future multi-baseline IR interferometry, both inclination and position angle will be self-consistently inferred from the data without the need to invoke extra information.

Priors for the inclination of the disk come from the kinematic modeling of the emission lines in the outflow cone[23,24]. Despite the differences in position angle from the different modeling approaches, the inclination of the system has been consistently found as 45°, with varying degrees of uncertainty. Accordingly, we sample inclinations from a normal distribution with a mean value of 45° and a standard deviation of ±10°.

**Modeling of the data**

*V- and K-band light curves:* In a first step, the *V*-band light curve was resampled to obtain a uniform coverage without seasonal gaps. The gaps were filled using a standard stochastic interpolation technique that makes use of the AGN's known variability pattern via the structure function[15,36]. The resampled *V*-band light curve was then used as the AGN variability pattern and put into a simplified radiative transfer model of a dust disk[18] to predict a model *K*-band light curve given $\tau_{in}$, the power law index $\alpha$ of the projected dust (brightness) distribution, the energy conversion efficiency $w_{eff}$, the sublimation temperature $T_{sub}$ of the dust, and the inclination i of the disk. Here, $w_{eff}$ represents the fraction of *K*-band light that reacts to the variability of the $V$-band[18]. In the framework of this study, it can be considered as a scaling factor of the amplitudes that does not have any influence on the time delay or shape of the variability pattern. As an important step in the process, we base our modeling on the relative light curve $g_i(t) = f_i(t)/\bar{f}_i(t) - 1$ $(i = V, K)$ rather than the light curve $f_i(t)$ to remove any dependencies on the dust covering factor. Here, $\bar{f}_i(t)$ is the mean flux of the respective band over the entire monitoring period.

**Extended Data Table 1 | Constraints on the inclination and position angle of the AGN structure from various observations.** The inferred distances for each individual observation do not include any uncertainties on the position angles and assume an inclination of 45°± 10°.

| method | inclination | position angle | approx. scale | reference | distance |
|---|---|---|---|---|---|
| optical continuum | ... | 91° | < 0."01 | continuum near H$\alpha$ line[25] | 20.4 ± 1.6 Mpc |
| polarimetry | ... | 92° ± 2° | < 0."01 | *UBVRI* polarimetry[34] | 20.5 ± 1.6 Mpc |
| kinematic modeling | 45° ± 5° | 35° | > 1" | bicone[23] | 15.7 ± 2.4 Mpc |
| of emission lines | 45° | 60° | < 2" | bicone[24] | 18.0 ± 1.8 Mpc |
| combined inference | 45° ± 10° | 75° ± 15° | ... | ... | $19.0^{+2.4}_{-2.6}$ Mpc |

To obtain the best representation of the resampled light curve, we implemented a fitting scheme based on the *idl mpfit* fitting package[37]. For that, the $K$-band light curve model fluxes have been extracted at the observed epochs. This model $K$-band fluxes have been compared to the observed $K$-band fluxes, and the squared residuals have been minimized by varying $\tau_{in}$, $\alpha$, and $w_{eff}$. The fitting scheme has been repeated for 1,250 randomly resampled $V$-band light curves to properly account for the uncertainties due to the finite sampling. In this process, $T_{sub}$ and $i$ have been treated as fixed parameter rather than actual model parameters since the light curves do not constrain these parameters. Instead we randomly draw a value from the prior distributions of $T_{sub}$ and $i$ for each resampled light curve that is fit. This accounts for the systematic uncertainty in these parameters.

Extended Data Figure 2 shows the 2-dimensional distribution of fitted $\alpha$ and $\tau_{in}$. The fully marginalized mean values and 68% confidence levels of these distributions are determined as $\tau_{in} = 36^{+9}_{-7}$ days and $\alpha =$ $-1.7^{+1.7}_{-1.2}$. Similarly, we found an efficiency factor of $w = 0.46^{+0.06}_{-0.07}$. These values are consistent with our previously published results[18].

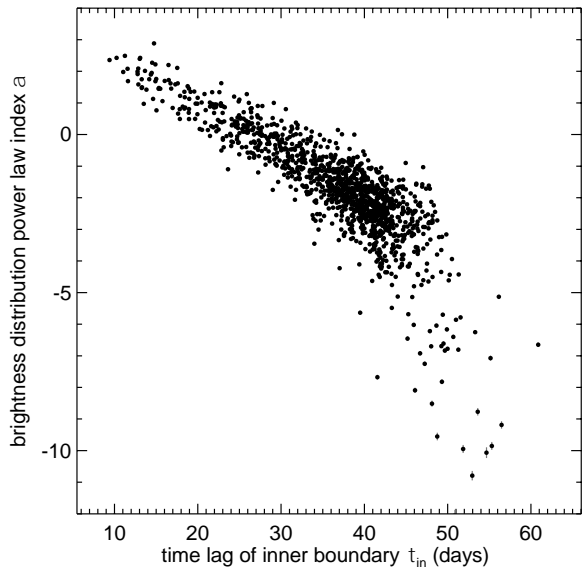

**Extended Data Figure 2 | Dependence of the time lag $\tau_{in}$ of the inner boundary on the brightness distribution power-law index α.** The black circles represent the fitted $\tau_{in}$ and α for each of the 1,250 Monte Carlo representations of the $V$-band light curve given the inclination and sublimation temperature distributions (68% fitting error levels mostly smaller than symbol size).

*K-band interferometry:* The light curve resampling and modeling provides us with 1,250 sets of parameters $(\tau_{in;j}, \alpha_j, T_{sub;j}, i_j)$. As shown in Fig. 2, the key element in obtaining a precise angular distance is to determine both time lag and IR interferometry size using the same set of parameters that define the brightness distribution of the object, in particular α. This parameter is correlated with both $\tau_{in}$ and $\rho_{in}$. While the α-dependence contributes a high uncertainty for constraints on $\tau_{in}$ and $\rho_{in}$ individually, it cancels as a source of uncertainty for the absolute distance, i.e. the ratio between both parameters, when determining $\tau_{in}$ and $\rho_{in}$ self-consistently given the inferred α.

From $\alpha_j$, $T_{sub;j}$, and $i_j$ we simulated model images of the $K$-band emission and fit the six observed squared visibilities with an angular size $\rho_{in}$ of the inner brightness distribution. This conversion from visibility to size depends on the orientation of the emitting disk since two-telescope long-baseline interferometry provides spatial information only along a specific position angle. Therefore, we randomly pick a disk orientation for each individual parameter set based on the prior distribution as constrained by observations (see previous section). This accounts for the systematic uncertainty in the orientation of the disk. We also correct for the contributions of the host galaxy and putative accretion disk to the $K$-band visibilities[12,14] and consider the respective uncertainties in our simulations. This additional Monte Carlo sampling has been repeated 200 times for each of the 1,250 initial sets of $(\alpha_j, T_{sub;j}, i_j)$ samples and each of the six independent interferometric measurements.

**Error budget and potential sources of systematic uncertainty**

The geometric distance measurement is based on a power-law parametrization for the brightness distribution in the $K$-band. Therefore, the uncertainties reflect the probability distribution *given this model* of 1,250 Monte Carlo samples to recover $\tau_{in}$ and α from the light curve and additional sampling to recover $\rho_{in}$ from the interferometric data. It accounts for uncertainties in the geometric parameters (disk inclination and position angle) as well as the statistical errors of the six interferometric observations and the uncertainties in their correction for host and putative accretion disk contributions as far as they are constrained by these data. The observations do not allow to distinguish between different parametrizations of the brightness distribution, which is an unknown uncertainty at this time. However, this parametrization was successful in simultaneously reproducing near- and mid-IR photometry and resolved interferometry of NGC 4151[17] and, therefore, seems a justified choice.

The error on the distance results from a combination of correlated and uncorrelated uncertainties in determining the reference time lag $\tau_{in}$, reference angular size $\rho_{in}$, and the brightness distribution power-law index $\alpha$. In Extended Data Figure 2, we show the distribution of $\tau_{in}$ and $\alpha$ values obtained from fitting the 1,250 random representations of the finite-sampled V-band light curve given inclination and $T_{sub}$ from the distributions discussed previously. The total scatter is dominated by the uncertainty in defining the "true" light curve from the limited number of epochs, with only a small contribution from the sampled range of inclinations. The relative contribution of this fitting to the total distance uncertainty can be estimated from the relative width of the distribution in $\tau_{in}$ given $\alpha$ and is approximately 6-7% (68% confidence level). Extended Data Figure 3 shows the distribution of $\rho_{in}$ for the 1,250 fitted $\alpha$ values, inclinations, and sublimation temperatures given the six interferometric observations. The gray error bars represent the additional error on $\rho_{in}$ from the uncertainty in position angle and in removing the host and putative accretion disk contributions, based on additional sampling of the interferometric data (see precious section). Similarly as before, we can estimate the relative contribution from $\rho_{in}$ from a combination of the scatter of the data points and the individual uncertainties as approximately 12% (68% confidence level). This combines to a total error in the distance of about 13.5% as obtained from the probability distribution function of $D_A$ shown in Fig. 2.

The major sources of systematic uncertainty are the object's inclination and orientation. The prior distributions of $i$ and the disk symmetry axis orientation are the dominant factors that contribute to the width of the $\tau$-$\rho$ correlation. The reason for this is that while the disk orientation influences the conversion of interferometric visibility to de-projected size, it does not have any influence on the time-delay signal. The inclination, on the other hand, affects both interferometry and reverberation signal. In interferometry, it causes a projection effect that decreases the observed size by approximately a factor of $\cos i \cdot \sin \theta$ as a function of the position angle offset $\theta$ from the disk symmetry axis. For reverberation mapping, however, it only broadens/smoothens the time lag signal symmetrically around the mean without a significant shift in $\tau$.

Although the dust covering factor cancels out when modeling relative light curves[18], we implicitly assume that the covering factor in the K-band is small. In the present model, the brightness distribution is a projected disk and, thus, geometrically thin. This simplifies the convolution of the light curves with the transfer function to a 2-dimensional problem. If there were significant amount of emission coming from material highly elevated above the disk, then we would obtain additional contributions to the transfer from the 3-dimensional space. Whether the presence of such emission would bias the results toward longer or shorter lags or a more compact or shallow brightness distribution, depends on the exact shape of the inner sublimation zone, which is unconstrained as of now. However, our assumption of a small covering factor is justified by the observed K-band covering factor of $c_K \sim 0.2$ as estimated from the mean fluxes of the V- and K-band light curves.

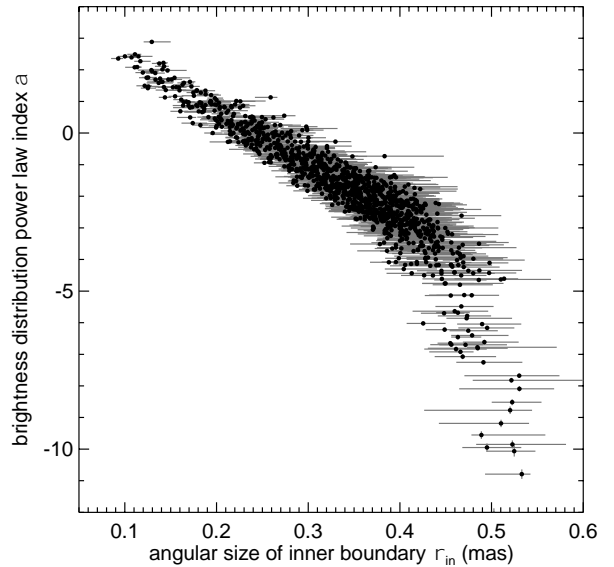

**Extended Data Figure 3 | Dependence of the angular size $\rho_{in}$ of the inner boundary on the brightness distribution power-law index $\alpha$.** The black circles represent the distribution of $\rho_{in}$ determined from the six interferometric data points for given $\alpha$, sublimation temperature, and inclination for each of the 1,250 Monte Carlo representations of the V-band light curve. The green error bars represent the additional uncertainties (68% confidence levels) from the combined statistical errors of the observations, the position angle of the emitting disk, as well as the corrections for host and putative accretion disk contributions.